\documentclass[aps,prl,twocolumn,amsmath,twoside,floatfix,superscriptaddress,preprintnumbers]{revtex4}
\preprint{CERN-PH-TH/2005-059}
\usepackage{graphicx}
\newcommand{\bea}{\begin{eqnarray}}
\newcommand{\beq}{\begin{equation}}
\newcommand{\eea}{\end{eqnarray}}
\newcommand{\eeq}{\end{equation}}

\newcommand{\nn}{\nonumber}
\newcommand{\Frac}[2]{\frac{\displaystyle{#1}}{\displaystyle{#2}}}
\newcommand{\lsim}{\raise0.3ex\hbox{$\;<$\kern-0.75em\raise-1.1ex\hbox{$\sim\;$}}}
\newcommand{\gsim}{\raise0.3ex\hbox{$\;>$\kern-0.75em\raise-1.1ex\hbox{$\sim\;$}}}
\newcommand{\bpm}{\begin{pmatrix}}
\newcommand{\epm}{\end{pmatrix}}
\newcommand{\eq}[1]{Eq.~(\ref{#1})}
\newcommand{\unity}{{\hbox{1\kern-.8mm l}}}
\newcommand{\Le}{{\rm L}}
\newcommand{\Ri}{{\rm R}}
\begin{document}

\title{Understanding the differences in neutrino and charged-fermion flavour 
structures}
\thanks{Dedicated to the memory of my father, M. Vives}
\author{O. Vives}
\affiliation{Theory Division, CERN, CH-1211, Geneva 23, Switzerland}

\begin{abstract}
We present a  mechanism to explain naturally and through a common flavour 
symmetry the mildly hierarchical neutrino masses with large mixings and
the hierarchical Yukawa matrices with small mixing angles. 
Although this mechanism is not linked to any particular flavour symmetry, 
it is particularly simple in the framework of a $SU(3)$ flavour symmetry. 
In this model, we
obtain exactly maximal atmospheric mixing, large although not maximal solar 
mixing, and a normal neutrino hierarchy. All neutrino parameters are basically 
fixed. For instance, we predict that the angle $\theta_{13}$ 
comes entirely from the charged lepton sector and, using the difference of 
the solar angle from maximality, we can fix the mass of the lightest neutrino.
\end{abstract}

\maketitle

%\twocolumngrid
In the last decade, oscillation experiments have confirmed that
neutrinos have non-vanishing mass and they do mix. However, it is 
not easy to understand the
measured values of neutrino masses and mixing angles \cite{fits}. These
parameters, specially when compared with quarks, are extremely
unusual.  On one side the quarks have masses of the order of
the electroweak scale, a strong intergenerational hierarchy and small
mixing angles. On the contrary, neutrino masses are much smaller than
the electroweak scale, they are not very hierarchical and the mixing 
angles are large.

The seesaw mechanism \cite{seesaw} can provide a very appealing
explanation for the neutrino mass scale. But even within this 
mechanism, the
intergenerational hierarchy and mixing angles remain a difficult
problem.  For example, in a $SO(10)$ Grand Unified Theory (GUT)
a full SM generation plus a right-handed neutrino are
included in the ${\bf 16}$ representation. Therefore, up quark 
and neutrino Dirac
Yukawa couplings are expected to be closely related.  
In these conditions, we could also expect a similar hierarchy of masses 
and small mixing angles in the neutrino sector.  On the contrary, 
the experimental results force us to obtain
near degeneracy and maximal mixing angles in the
left-handed Majorana neutrino mass matrix from a strong hierarchy and
small mixing angles in the Yukawa matrices.
In the literature we can see a large variety of models that succeed to do 
this \cite{models}.  However, these models
are not completely satisfactory. In some cases, as we find for instance 
in $U(1)$ flavour models or in some $SU(5)$ GUTs, large mixings come
from the the charged lepton sector, but here it is not straightforward
to guarantee maximal atmospheric and large solar angles together with
correct mass differences. In pure $SO(10)$ or non-Abelian flavour
symmetries, we usually have symmetric Dirac neutrino Yukawa
matrices. Then, we have to obtain large mixings from small
Cabibbo--Kobayashi--Maskawa (CKM) mixings. This is usually obtained at 
the cost of a certain tuning of the parameters. In any case, it is evident 
that the
(flavour) symmetry structure of the low-energy effective neutrino mass
matrix is radically different from the very hierarchical structure of
charged leptons and quarks. This is the source of all our problems
when trying to accommodate simultaneously quarks and leptons in the
same flavour model. So far there is no convincing symmetry reason to
ensure large neutrino mixings together with small quark mixing angles.

In this paper we present a very simple mechanism that allows us to
obtain naturally correct neutrino masses and large mixing angles from,
for instance, very hierarchical Yukawa matrices with small mixing. All
this is obtained with all the fermion mass matrices including
right-handed Majorana matrices sharing basically the same flavour
structure. The only small differences are precisely due to the
Majorana nature of the neutrino mass matrix. In the following,
we briefly present the main
ideas of this mechanism and then we will discus some possible
realization of this mechanism in flavour models.
    
In the literature, it is known that to obtain the experimental values 
of the neutrino mass differences through the seesaw mechanism, the 
hierarchy in the right-handed Majorana matrix must be approximately 
the square of the hierarchy in the Yukawas \cite{nuhierarchy,smirnov}.  
The simplest possible realization of this idea would be to take 
$M_\Ri = M_{\rm X} (Y_{\nu}^T Y_{\nu})$. From this structure 
we would obtain exact neutrino degeneracy from any Dirac neutrino 
Yukawa matrices \cite{smirnov}. 
Although this goes in the right direction, it is not enough: 
neutrinos {\it are not} degenerate, and large mixings are present
in nature. However, continuing along the same lines, it is straightforward to 
obtain the required right-handed Majorana matrix from the seesaw formula 
itself:
\bea
\label{invseesaw}
M_\Ri = v_2^2~Y_{\nu}^T\cdot\left(\chi_\Le\right)^{-1}\cdot Y_{\nu}
\eea
This equation suggests that the structure of the
right-handed Majorana mass matrix is proportional to $Y_{\nu}$ on the left 
and the right, and that these matrices are connected by
a third flavour matrix, which is simply the inverse of the
low-energy effective neutrino mass matrix. This is the only possible 
solution once we fix the neutrino Yukawa matrices.  This expression
is analogous to the $M_\Ri$ we would obtain in the double seesaw mechanism 
\cite{mohapatra-valle}. Although \eq{invseesaw} does not mean that
we must go through a double seesaw to obtain $M_\Ri$ with this structure, 
we think it is really worth trying to construct a
right-handed Majorana matrix exactly along these lines.

Using this form it is easy to see that $M_\Ri$ has the following features:
i) The texture of $M_\Ri$ is still diagonalized by similar rotations to 
the rotations diagonalizing the Yukawa matrices, and 
the hierarchy in $M_\Ri$ is approximately the square of the hierarchy 
in $Y_\nu$.    
ii) Even in this situation, the structure
of the left-handed neutrino Majorana matrix is {\bf basically decoupled} from 
the texture of the Yukawa matrices of neutrinos and charged fermions.
iii) The texture of the effective left-handed neutrino Majorana mass matrix 
is only  determined by the structure of the connecting matrix $\chi_\Le^{-1}$.

This strategy was originally proposed by A.~Y.~Smirnov in Ref.~\cite{smirnov}. 
In this letter we justify these features in the context of a flavour symmetry 
and show that this mechanism is particularly efficient in a
$SU(3)$ flavour theory. As we will show, in this way we can expect $M_\Ri$ 
and $Y_\nu$ to be determined by the {\bf same flavour symmetry}.

It is very interesting to analyse what the required structure of
this connecting matrix $\chi_\Le^{-1}$ is. Let us assume, that we have 
bimaximal neutrino mixings. The mixing matrix $U$ then is 
\bea
\label{Umix}
U=\bpm \frac{1}{\sqrt{2}} & \frac{1}{\sqrt{2}} & 0 \\-\frac{1}{2} &\frac{1}{2}
&\frac{1}{\sqrt{2}}
\\\frac{1}{2} &-\frac{1}{2}& \frac{1}{\sqrt{2}} \epm.
\eea
Regarding the neutrino masses, we only know two mass differences but there is
absolutely no reason to believe that one of the masses is exactly zero and 
thus we leave them free for the moment. Then, $\chi_\Le = U^*\cdot 
\mbox{Diag}\left(m_1,m_2,m_3\right) \cdot U^\dagger$,
and trivially the inverse of this matrix is $\chi_\Le^{-1} = U\cdot 
\mbox{Diag}\left(\frac{1}{m_1},\frac{1}{m_2},\frac{1}{m_3}\right)
\cdot U^T$. Therefore,
\bea
\label{connect1}
\chi_\Le^{-1} =&  \Frac{1}{m_3} \bpm 0&0&0\\
0&\frac{1}{2}&\frac{1}{2}\\
0&\frac{1}{2}&\frac{1}{2}\epm
 + \Frac{1}{m_2}  
\bpm \frac{1}{2}&\frac{1}{2 \sqrt{2}}&-\frac{1}{2 \sqrt{2}}\\
\frac{1}{2 \sqrt{2}}&\frac{1}{4}&-\frac{1}{4}\\
-\frac{1}{2 \sqrt{2}}&-\frac{1}{4}&\frac{1}{4}\epm &\nn \\& 
+ \Frac{1}{m_1}  \bpm \frac{1}{2}&-\frac{1}{2 \sqrt{2}}&\frac{1}{2 \sqrt{2}}\\
-\frac{1}{2 \sqrt{2}}&\frac{1}{4}&-\frac{1}{4}\\
\frac{1}{2 \sqrt{2}}&-\frac{1}{4}&\frac{1}{4} \epm.  &
\eea
This structure is indeed very simple and, given that all the elements
in the matrices are of the same order, the only possibly small
parameters that can order the different contributions are the neutrino 
mass eigenvalues themselves. Maximal atmospheric
mixing is given by the first contribution in \eq{connect1}
and in fact this structure is {\it precisely} one of the main ingredients of
$SU(3)$ flavour models. Although,
in principle, it could be possible to reproduce this structure in
other flavour models, we think it is a hint in favour of these models
and we will exploit this relation in the following.
 
The main question now is whether it is possible in a complete model to
obtain the structure given in \eq{invseesaw} or, more exactly, in 
\eq{connect1}. To reach this goal we
have to force the right-handed neutrinos to couple ``only'' through
the combination $(\nu^c_\Ri)_j (Y_\nu)^{ij}$. This combination carries
exactly opposite charges to $H \nu_\Le$ if we
assume that the Dirac Yukawa coupling is allowed. This
means that it carries minus one unit of lepton number plus the flavour
charge that would be saturated by the left-handed neutrino. Now we
have to remember that the Majorana matrix has still another
fundamental feature: it violates lepton number
(or ($B$--$L$)) by two units.  This means that to obtain a
Majorana matrix from a product of Yukawa matrices we would need to add
a vacuum-expectation-value (vev) breaking lepton number. 

We can think of a scalar field $\lambda$, with $L=1$ and
charged under the flavour symmetry saturating the quantum numbers of
$(\nu^c_\Ri)_j (Y_\nu)^{ij}$, which gets a vev. In this way we would need
a new mediator $S$, singlet under the SM, flavour and ($B$--$L$)
symmetries, with a relatively large mass to connect again with
$(\nu^c_\Ri)_i (Y_\nu^T)^{ij}$. Clearly, this construction  is 
allowed by all the symmetries. This would be a double seesaw mechanism for 
$((\nu_\Le)_i,(\nu^c_\Ri)_i,S)$ \cite{mohapatra-valle}. In fact it is 
simpler to discuss the physics in terms of the double seesaw structure. 
The ``neutrino'' mass matrix here is 
\bea
\label{2seesaw}
\bpm (\nu_\Le)_i&(\nu^c_\Ri)_j&S \epm \cdot \bpm 0 & Y^{ij} v &0 \\ 
Y^{ij} v & 0& 
Y^{kj}\cdot \langle \lambda \rangle_k \\ 0 &  \langle \lambda \rangle^T_k
\cdot Y^{kj} & M_S \epm \cdot \bpm (\nu_\Le)_i\\(\nu^c_\Ri)_j\\S \epm
\eea
with $v$ the electroweak Higgs vev.
First, we notice that this structure can not reproduce the correct
spectrum. It gives Majorana mass to only one right-handed
and one left-handed neutrino and the remaining neutrinos get Dirac
masses of order the electroweak scale. This can be solved by adding 
other fields coupling to $(\nu^c_\Ri)_j (Y_\nu)^{ij}$ in a similar 
way to generate Majorana masses to the other neutrinos. In fact, we 
need two other singlets $S^\prime$ and $\hat S$. However, the role of 
the flavon vev $\lambda$ in the equation above can be played by a 
combination of other flavon fields and $\lambda$, without introducing new 
$B-L$ violating vevs as we show in a $SU(3)$ example below. 
Equation~(\ref{2seesaw}) (with additional singlets) is the structure 
we need, but clearly there 
are several conditions our model has to fulfill to obtain this matrix: 
\begin{enumerate}
\item We must make sure that the same Yukawa matrix enters both the 
$(1,2)$ and $(2,3)$ blocks.  
\item We should guarantee that there is no direct mixing between $\nu_L$ 
and $S$, i.e. the $(1,3)$ entry is vanishing or sufficiently small.
\item There could be other operators with a structure also
allowed by our symmetries contributing directly to the 
(2,2) block of \eq{2seesaw}, which should be sufficiently small. 
\end{enumerate}
The first condition is automatically satisfied if $\lambda$ behaves in
the same way as $\nu_L$ under the flavour symmetries.  Points 2 and 3
are also solved if $\lambda$ is the only field breaking lepton number
and has $L=1$. However, in the context of a Supersymmetric model, we
would expect other fields $\bar\lambda$ with opposite charges and
equal vev as $\lambda$ to maintain D-flatness. Then point 2, in the context
of a Froggatt--Nielsen mechanism \cite{froggatt}, and taking
into account $SU(2)_L$ and ($B$--$L$) symmetries, means that the coupling
$\frac{1}{M_{\rm fl}}(\nu_L)_i f^{i}_j\bar \lambda^j H S$, should be
sufficiently small. Where $f^{i}_j$ is a Yukawa-like function of flavon
vevs and $M_{\rm fl}$ one of the flavour mediators that enter the
Froggatt--Nielsen graph. Its contribution to left-handed neutrino
masses would be $ m_\nu^\prime = \frac{v^2}{M_S}~f^{i}_k\langle \tilde
\lambda \rangle^k f^{j}_l\langle \tilde \lambda
\rangle^l\frac{v_{B-L}^2}{M_{\rm fl}^2}$, with $\langle \bar \lambda \rangle
=\langle \lambda\rangle=
v_{B-L}\langle\tilde \lambda\rangle$ and $v$ the Higgs $SU(2)$-breaking vev.  
This has to be compared to the double seesaw contribution, which would be
simply $\frac{v^2 M_S}{v_{B-L}^2}\langle \tilde \lambda \rangle_i
\langle \tilde \lambda \rangle_j$. Given that at least some entries
in the $f_{j}^i$ matrices can be of order 1, we do not consider any
additional suppression from them. Then the condition that this
contribution is small is simply 
\bea
\label{cond1}
\frac{v_{B-L}^2}{M_S M_{\rm fl}^2} \ll \frac{M_S}{v_{B-L}^2}
\Rightarrow \frac{v_{B-L}^4}{M_S^2 M_{\rm fl}^2} \ll 1.
\eea
Point 3 is satisfied in our case by construction because the only
field that breaks ($B$--$L$) is $\lambda$ (and possibly $\bar
\lambda$). $\lambda$ saturates the charges of the combination
$(\nu^c_\Ri)_j (Y_\nu)^{ij}$ and therefore couples to our singlet
field as shown above. The (2,2) block could be directly filled, for 
instance, in the presence of a
scalar vev with $L=2$ and some flavour structure connecting 
$(\nu^c_\Ri)_i(\nu^c_\Ri)_j$. In this case, all the mediator fields in the
Froggatt-Nielsen diagram determining the flavour structure would be
charged under the flavour and ($B$--$L$) symmetries. If $M_{\rm fl}$
is the mass of these mediators, our strategy to suppress these
contributions would be to require, $M_S \ll M_{\rm fl}$. As said above, in a
Supersymmetric model Eq.~(\ref{cond1}) is relevant. To fulfill this condition 
is enough to take the scale of ($B$--$L$) breaking much below the scale 
of flavour breaking. In the presence of additional scalar vevs, $M_S$ 
can also be taken lower than $M_{\rm fl}$. For instance, we can have 
the flavour
symmetry broken at the Planck or GUT scale, while the scale of
($B$--$L$) breaking and $M_S$ are both around the scale of the
right-handed neutrinos, several orders of magnitude below.  As $S$ is
a singlet under the gauge group, there is no problem to have its mass
at this scale.  In this way, we obtain the structure in \eq{2seesaw}.
Notice that, although we have discussed these conditions in
terms of $S$, they apply in the same way to the additional singlets
$S^\prime$ and $\hat S$.

To demonstrate the applicability of this mechanism we present a full
realization in a realistic $SU(3)$ (Supersymmetric) flavour model 
\cite{SU(3)}. 
The basic features of this symmetry are the following. All left-handed 
fermions ($\psi_i$ and $\psi^c_i$) are triplets under $SU(3)_{\rm fl}$. To
allow for the spontaneous symmetry breaking of $SU(3)$ it is necessary to
add several new scalar fields, which are either triplets ($\overline{\theta}%
_{3}$, $\overline{\theta}_{23}$) or antitriplets ($\theta_{3}$, 
$\theta_{23}$). We assume that $SU(3)_{\rm fl}$ is broken in two
steps. The first step occurs when $\theta_3$ and $\bar \theta_{3}$ get
a large vev of the order of the mediator scale, $M_{\rm fl}$, breaking $SU(3)$ 
to $SU(2)$. Subsequently a smaller vev of 
$\theta_{23}$ and $\bar \theta_{23}$, of order  $\lambda_c$, the Cabibbo angle 
in the down sector and $\lambda_c^2$ in the up sector, breaks the
remaining symmetry. Moreover in the minimization of the scalar potential 
the fields $\theta_{23}$ and $\bar \theta_{23}$ get equal 
vevs in the second and third components. Notice that, although this condition
was originally introduced to obtain $V_{cb}\sim m_s/m_b$ in these models, 
this is precisely 
the required feature to reproduce maximal atmospheric mixing from 
\eq{connect1}. Effective Yukawa
couplings are obtained through the Froggatt--Nielsen mechanism 
\cite{froggatt}. 
Third-generation Yukawa couplings are generated by
$\theta _{3}^{i}\theta _{3}^{j}$ while couplings in the 2--3 block of
the Yukawa matrix are always given by $\theta _{23}^{i}\theta
_{23}^{j}$. The couplings in the first row and column of the Yukawa
matrix are given by $\epsilon ^{ikl}\overline{\theta
}_{23,k}\overline{\theta }_{3,l} \theta _{23}^{j}\left( \theta
_{23},\overline{\theta_{3}}\right)$, where $\left( \theta
_{23},\overline{\theta_{3}}\right) \equiv \theta
_{23,i}~\overline{\theta_{3}}^i$. Additional global symmetries are usually
imposed to forbid unwanted terms in the effective
superpotential, such as a mixed $\theta_{3}^{i}\theta _{23}^{j}$ term.

In this framework, we add a new $SU(3)$ triplet $\lambda$ with $L=1$
that gets a vev breaking lepton number. Therefore, a coupling
$(\nu^c_\Ri)_i (Y_\nu^T)^{ij} (\lambda)_j~ S$ is automatically allowed
by the symmetries. We must now ensure that $\langle\lambda\rangle^T =
(1/\sqrt{2},-1/2,1/2)$ to reproduce the phenomenologically observed
mixings. We do not want to present a full minimization of the scalar
potential of the model here, but in fact to obtain this vev is very
simple in this model. First we need another field $\overline{ \lambda}$,
antitriplet of $SU(3)$ to ensure D flatness \cite{SU(3)}. The required vev
is obtained minimising the following F terms in the scalar potential 
together with D-flatness: 
$F_1=(\theta_{23}, \lambda)$,
$F_2=(\overline{ \lambda},\overline{\theta_{23}})$ and
$F_3=(\overline{ \lambda},\lambda)$ \cite{SU(3)}.  Then, this contribution 
would give rise to an effective coupling $1/M_S (\nu^c_\Ri)_i (Y_\nu^T)^{ik} 
\langle \lambda_k \rangle \langle \lambda_l \rangle^T (Y_\nu)^{lj} 
(\nu^c_\Ri)_j$. This
means that the third term in \eq{connect1} is given exactly by
$\langle \lambda \rangle \langle \lambda \rangle^T$. We still need the
other two contributions in \eq{connect1} to give Majorana masses to 
the three generations. In our model, we have other $SU(3)$ triplets that
may also couple to $(\nu^c_\Ri)_i (Y_\nu^T)^{ij}$. For instance
requiring lepton number conservation, we could also have
$(\nu^c_\Ri)_i (Y_\nu^T)^{ij} (\overline{ \theta_{23}})_j
(\theta_3,\lambda)~ S^\prime$. This gives rise to the first matrix in
\eq{connect1} as $\langle (\overline{ \theta_{23}}) \rangle \langle
(\overline{ \theta_{23}}) \rangle^T$ with a coefficient of, at least,
$(\theta_3,\lambda)^2$. Notice that we need a new singlet $S^\prime$,
distinguished from $S$ by a $Z_2$ symmetry, to avoid
an interference $\langle \lambda \rangle\langle
(\overline{ \theta_{23}}) \rangle^T$ that would disturb \eq{connect1}.

The easiest way to generate the second matrix  would be to introduce a new 
flavon field $\lambda^\prime$ carrying lepton number and having a vev 
orthogonal to $\lambda$ and $\theta_{23}$ together with a new singlet 
$\hat S$. Still, it is possible to do it only with the 
flavon vevs already present in our $SU(3)$ flavour theory without any other
($B$--$L$)-violating vev.
The trick to do this is to notice that rows and columns of the
matrices in \eq{connect1} are orthogonal if they do not
belong to the same matrix. So, if we want to preserve maximal atmospheric 
mixing and
we do not mind perturbing slightly maximal solar mixing, it is sufficient
to use another vector orthogonal to $\overline{\theta_{23}}$, even
though it is not orthogonal to $\lambda$. Only with the
fields already present in the model, we have the combination
$(\theta_3,\lambda) \epsilon_{ikl} \theta_{23}^k \theta_3^l$, coupling
$(\nu^c_\Ri)_j (Y_\nu)^{ij} $ to $\hat S$, 
which is a vector in the direction $(1,0,0)$. 
Therefore this term would provide a third matrix with a single non-vanishing 
entry in the $(1,1)$ element. As we will see, this matrix does not change 
maximal atmospheric mixing. Defining $\tilde m_2$ as 
the mass scale associated to this last matrix, $\tilde m_3$, $\tilde m_1$ 
the mass scales associated with the first and third matrices in 
\eq{connect1} respectively, and using $x \equiv
\tilde  m_1/\tilde m_2$ and $y \equiv \tilde m_1/\tilde  m_3$ we have:
\bea
\label{connectSU3}
\chi_\nu^{-1} = \frac{1}{\tilde m_1}
\bpm \frac{1}{2} + x & -\frac{1}{2\sqrt{2}}
&\frac{1}{2\sqrt{2}}\\
-\frac{1}{2\sqrt{2}}&\frac{1}{4}+
\frac{y}{2}&
-\frac{1}{4}+\frac{y}{2}\\
 \frac{1}{2\sqrt{2}} &-\frac{1}{4}+
\frac{y}{2}&\frac{1}{4}+\frac{y}{2} \epm. 
\eea
Now, we can use the fact that the rotation in the $(2,3)$ sector is still 
maximal and undo this rotation:
\bea
R_{23}\left(\frac{\pi}{4}\right)\cdot\chi_\nu^{-1}\cdot R_{23}^T
\left(\frac{\pi}{4}\right)=\frac{1}{\tilde m_1}
\bpm
\frac{1}{2} +x & -\frac{1}{2}&0\\
  -\frac{1}{2}&\frac{1}{2}&0\\
0&0&y\epm.
\eea
From there, it is straightforward to diagonalize 
the $(1,2)$ submatrix
and we obtain, to order $x^2$, two eigenstates of masses $m_{1}^{-1}= 
\tilde m_1^{-1} \left(1 + x/2 + x^2/4\right)$ and $m_{2}^{-1} 
= \tilde m_1^{-1} \left(x/2- x^2/4\right)$ and 
$\sin \theta_{12} = (1 - x/2 - x^2/4)/\sqrt{2}$.  Therefore $x$ must be 
relatively small to preserve large solar mixing angle. The solar and 
atmospheric 
mass differences \cite{fits} are trivially obtained from the absolute 
scale and, using $m_3=\tilde m_3$, from the relation $m_2/m_3 \simeq 2 y/x 
\simeq 1/6$, which fixes $y = x/12$.
Now, taking also into account the charged lepton rotation, the difference
from maximal solar mixing fixes the value of $x$; therefore, we give
a prediction for the mass of the lightest neutrino. Furthermore, given
the experimental values for neutrino mass differences and angles, in
this example we can only obtain normal neutrino hierarchy. Moreover 
before the rotation from the charged lepton sector is taken into account
the $\theta_{13}$ angle is naturally zero, if we want to 
maintain maximal atmospheric mixing. In this model, we expect a 
non-negligible $\theta_{12}^l$ rotation in the charged
lepton sector, which would give rise to a $\theta_{13} =
\theta_{12}^l/\sqrt{2}$ (in simple models with a Georgi--Jarlskog factor  
\cite{georgi} we would expect $\theta_{12}^l = \lambda_c/3$, but other 
relations are possible \cite{antusch}).
Similarly right-handed neutrino masses can also be fixed, although they 
depend on the Dirac neutrino Yukawas, which we did not have to specify so far.
Full phenomenological details will be presented elsewhere.
Finally we have to ensure that these
contributions are really the dominant flavon contributions to neutrino masses
through this mechanism. For instance, a contribution proportional to
$\overline{\theta_3}$ would spoil our predictions if it were not
sufficiently suppressed but this can easily be done for instance with the
help of a discrete symmetry forbidding this term.

We have constructed a mechanism to explain naturally neutrino masses and
mixing angles consistently with charged fermion Yukawas. This mechanism is 
particularly simple in the framework of an $SU(3)$ flavour theory. 
In this model we have been able to obtain maximal atmospheric mixing, large 
although not maximal solar mixing, and correct mass differences. As a bonus, we
predict that the angle $\theta_{13}$ comes entirely from the charged lepton 
sector and the lightest neutrino mass is fixed in terms of the known 
parameters.

\noindent
\textit{Acknowledgements:}
I acknowledge support from the RTN European project MRTN-CT-2004-503369 and 
from the Spanish  MCYT FPA2002-00612.
I thank J. Bernabeu, R. Barbieri, F.J. Botella, G. Giudice, A. Romanino, 
G.G. Ross and A. Santamaria for discussions. I am indebted to A. Smirnov for
helpful and clarifying remarks and the comparison with Ref.~\cite{screening}.

\end{document}